\documentclass[%
aps,
prb,
twocolumn,
groupedaddress,
]{revtex4-2}

\usepackage[pdftex]{graphicx}
\usepackage{bm}
\usepackage{amsmath,amssymb,amsfonts}
\usepackage{braket}

\begin{document}

\title{ Lifetime of edge modes at rough surfaces of chiral
  superconductors }

\author{Seiji Higashitani}
%\email[]{ultp@hiroshima-u.ac.jp}
\author{Gota Sato}
\author{Yasushi Nagato}
\affiliation{
  Graduate School of Advanced Science and Engineering,
  Hiroshima University,
  1-7-1 Kagamiyama, Higashi-Hiroshima City, Hiroshima, 739-8521, Japan
}

\date{\today}

\begin{abstract}
  We study the effect of diffuse surface scattering on the edge modes
  in two-dimensional chiral superconductors with time-reversal
  symmetry-breaking Cooper pairs, each carrying angular momentum
  $m\hbar$ ($m = 1,2,3,\cdots$). To elucidate the diffuse scattering
  effect, we formulate the inverse lifetime $\Gamma = \hbar/\tau$
  corresponding to the broadening of the surface density of states
  (SDOS) for the edge mode. This derivation uses random
  S-matrix theory, which allows us to describe the surface effect in a
  unified way from the specular to the diffuse limit within the
  quasiclassical theory framework of superconductivity.  We
  find that $\Gamma$ in the chiral states with $m \geq 2$ is 
  larger than that in the chiral $p$-wave state ($m = 1$) because of the multiple edge mode branches in former superconducting
  states (the number of which 
  equals $m$).  Diffuse scattering between the different branches
  causes a significant broadening of SDOS. In contrast, the edge
  mode in the chiral $p$-wave state is robust to diffuse
  scattering because only a single edge mode branch exists. We also
  discuss SDOS at the diffuse limit, where the description in terms of
  $\Gamma$ is not useful. The diffuse scattering effect on SDOS can
  be understood qualitatively in terms of destructive
  interference analogous to that caused by impurity scattering in
  unconventional superconductors.
  
\end{abstract}

\maketitle

\section{Introduction \label{sec:intro}}

Since the discovery of superfluid $^3$He, Cooper pair condensates with
time-reversal symmetry-breaking chiral order have been of continuing
interest in condensed matter physics. It is well established that
spin-triplet $p$-wave pairing is realized in superfluid $^3$He and the
superfluid $^3$He-A phase is in the chiral $p$-wave pairing state
\cite{Vollhardt_Book1990superfluid}.
Triplet pairing analogous to
superfluid $^3$He was proposed for the quasi-two-dimensional (2D)
superconductivity in Sr$_2$RuO$_4$ shortly after its discovery
\cite{Maeno_Nature1994,Rice_IOP1995}.
Subsequent
experiments supported the 2D chiral $p$-wave state among the proposed
candidates
\cite{Maeno_JPSJ2012}.
This has led to numerous
theoretical investigations into 2D chiral $p$-wave superconductivity (note,
the pairing symmetry of Sr$_2$RuO$_4$ has recently been
reexamined and is currently under discussion
\cite{Maeno_JPSJ2024}).
The 2D chiral $p$-wave state belongs to the
group of pairing states characterized by the gap function
$\propto (\hat{k}_x \pm i\hat{k}_y)^m$, where
$\hat{k} = (\hat{k}_x, \hat{k}_y)$ is the unit vector along the Fermi
wavevector and $m = 1,2,3,\cdots$. The case of $m=1$ corresponds to
the chiral $p$-wave state.

The surface state of
unconventional (non-$s$-wave) superconductors differs 
from that of the bulk state.  The surface physics in the chiral superfluid
and superconductors are highlighted by the gapless edge
mode, which induces an edge current along the surface
\cite{Matsumoto_JPSJ1999,Stone_PRB2004,Sauls_PRB2011,
  Tsutsumi_PRB2012, Mizushima_JPSJ2016,
  Huang_PRB2014,Tada_PRL2015,Wang_PRB2018,Sugiyama_JPSJ2020,
  Holmvall_PRB2023,Ashby_PRB2009,Nagato_JPSJ2011,Bakurskiy_PRB2014,
  Lederer_PRB2014,Suzuki_PRB2016,Bakurskiy_IOP2017,Suzuki_PRB2023}.
In the chiral $p$-wave state, the edge current carried by the
edge mode is sizable at low temperatures. According to the
quasiclassical theory of superconductivity, the corresponding mass
current at absolute zero is estimated to be $n\hbar/2$ with $n$ being
the number density of fermions
\cite{Stone_PRB2004,Sauls_PRB2011,Tsutsumi_PRB2012,Mizushima_JPSJ2016}.
However, the edge mass
current vanishes in higher-order chiral states, such as those with
$d$-wave ($m = 2$) and $f$-wave ($m = 3$) orbital symmetry
\cite{Huang_PRB2014,Tada_PRL2015,Wang_PRB2018,Sugiyama_JPSJ2020}.
This contrasting result is because, unlike the $p$-wave state,
the higher-order states have multiple edge-mode branches and their
contributions to the mass current cancel out. Moreover, the orbital symmetry-dependent feature of chiral
superconductivity manifests in the edge current density at a rough
surface.
In real systems, surfaces inevitably have microscopic
irregularities that are sufficient for quasiparticles to diffusely
scatter and suppress the edge current
density. Diffuse scattering effects on the chiral superconductivity
have been extensively studied as an important factor affecting the
observability of edge currents
\cite{Ashby_PRB2009,Nagato_JPSJ2011,Bakurskiy_PRB2014,Lederer_PRB2014,
  Suzuki_PRB2016,Bakurskiy_IOP2017,Suzuki_PRB2023}.
According to the numerical
analysis of Suzuki and Asano
\cite{Suzuki_PRB2016}, the influence of diffuse scattering is
weak for the chiral $p$- and $d$-wave states but is substantial
for the chiral $f$-wave state that the edge current density is almost
suppressed to zero.

In this work, we focus on the broadening of the surface density of
states (SDOS) for the edge mode caused by diffuse scattering.
In particular, we ask the following questions that have not been
addressed in detail: what is the main scattering process that 
broaden SDOS of the edge mode
and to
what extent does the magnitude of the broadening differ among chiral
superconductors with different orbital symmetry?  To address these
questions, we formulate SDOS in terms of a surface self-energy that
contains information about the lifetime of the edge mode. We use the random S-matrix theory, which provides a closed set of
quasiclassical equations to describe the rough surface effect on
superconductivity
\cite{Nagato_JLTP1996,Nagato_JLTP1998,Nagai_JPSJ2008,Miyawaki_PRB2018}.

The remaining paper is organized as follows.  In Section\ \ref{sec:model}, we
describe our theoretical model. In Section\ \ref{sec:Edge mode at
  specular surface}, we briefly review the edge mode formed at a
specular surface of the 2D chiral superconductors and classify the
edge mode into two.
This classification is important for understanding the effects of diffuse scattering on the edge modes.
In Section\ \ref{sec:Method}, the
general formula for SDOS is derived based on the random S-matrix
theory and some numerical results of SDOS are presented. In Section\
\ref{sec:SDOS of the edge mode}, we first introduce the surface self-energy
for the edge mode, and then derive an approximate formula for subgap
SDOS, which applies to the case where the diffuse scattering
probability is sufficiently small and the edge mode is well
defined. The approximate formula is used to define the inverse lifetime
$\Gamma = \hbar/\tau$ of the edge mode.
Additionally, we briefly discuss SDOS in
the diffuse limit, where the scattering is completely diffuse. The
final section presents the conclusion.

\section{Model}
\label{sec:model}

\begin{figure}
  \includegraphics[scale=1.1]{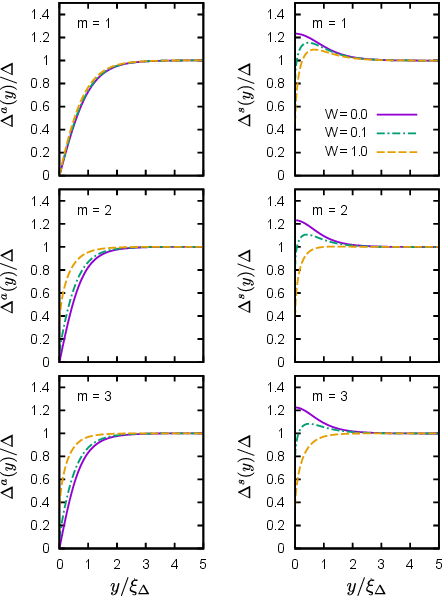}
  \caption{Spatial dependence of $\Delta^a(y)$ (left panels) and
    $\Delta^s(y)$ (right panels) for $m = 1,2,3$ at
    $W = 0.0, 0.1, 1.0$.  The results are calculated at 
    $T = 0.2 T_c$, where $T_c$ is the transition temperature. The
    distance $y$ from the surface is scaled by the coherence length
    $\xi_\Delta = \hbar v_F / \Delta$.}
    \label{fig:Delta-y}
\end{figure}

We consider a semi-infinite 2D superconductor occupying the $y > 0$
space and with a rough surface along the $x$-axis. The surface is
assumed to be macroscopically flat but with atomic-scale
irregularity. The superconducting state is characterized by the
following chiral gap function in the bulk:
\begin{equation}
  \Delta(y \to \infty, \hat{k}) = \Delta e^{im\phi},
\end{equation}
where $\Delta$ is the gap amplitude with a positive real value and
$\hat{k} = (\hat{k}_x, \hat{k}_y) = (\cos\phi, \sin\phi)$.  The
orbital symmetry of the gap function is classified by an integer $m$.
For example, $m = 1,2,3$ corresponds to the $p$-, $d$-, and $f$-wave
superconductors, respectively. The bulk gap function, defined as a spin
space matrix, has the form $\Delta e^{im\phi} i\sigma_2$ for the
singlet superconductors with even $m$ and $\Delta e^{im\phi} \sigma_1$
for the triplet superconductors with odd $m$, where $\sigma_i$ ($i=1,2,3$) are
Pauli matrices.

Surface scattering is known to cause a spatial variation of the gap
function.
Consequently, the gap function takes the following form:
\begin{equation}
  \Delta(y,\hat{k}) = \Delta^s(y,\phi) + i \Delta^a(y,\phi),
  \label{eq:Delta-def-tmp}
\end{equation}
where $\Delta^{s}$ and $\Delta^{a}$ denote the symmetric and
antisymmetric components with respect to the transformation
$\phi \to -\phi$, respectively, i.e.,
\begin{align}
  \Delta^s(y,\phi) &= \Delta^s(y)\cos(m\phi), \\
  \Delta^a(y,\phi) &= \Delta^a(y)\sin(m\phi).
\end{align}
$\Delta^s(y)$ and $\Delta^a(y)$ asymptotically
approach the bulk value $\Delta$ as $y \to \infty$.  Typical examples
of the self-consistently determined spatial dependence of
$\Delta^s(y)$ and $\Delta^a(y)$ are shown in Fig.\ \ref{fig:Delta-y},
where the parameter $W$ controls the diffuse scattering
probability [see Eq.\ \eqref{eq:gamma-W} and below for
details]. Similar results for the rough surface effect on the
self-consistent gap functions in the chiral $p$-, $d$-, and $f$-states
can be found in Refs.\
\onlinecite{Wang_PRB2018,Suzuki_PRB2016}.

When analyzing the boundary problem in the semi-infinite
superconductor, it is convenient to restrict the domain of $\phi$ to
$[0, \pi]$ and specify the position on the Fermi surface by the unit
vector $\hat{k}_\alpha = (\cos\phi, \alpha \sin\phi)$ with
$\sin\phi > 0$ and $\alpha = \pm$.  Accordingly, Eq.\
\eqref{eq:Delta-def-tmp} is expressed as
\begin{equation}
  \Delta(y,\hat{k}_\alpha) = \Delta^s(y,\phi) + \alpha i \Delta^a(y,\phi)
  \equiv \Delta_\alpha(y,\phi).
  \label{eq:Delta-def}
\end{equation}
The gap function $\Delta_\pm(y,\phi)$ corresponds to the pair
potential felt by a quasiparticle on the Fermi surface with a wave
function $\propto e^{ik_\parallel x \pm ik_\perp y}$, where
$k_\parallel = k_F \cos\phi$ and $k_\perp = k_F \sin\phi$.

\section{Edge mode at specular surface}
\label{sec:Edge mode at specular surface}

When the surface is specular, the quasiparticle states in the above
model system can be described by the wave function
\begin{equation}
  \psi(x,y)
  = e^{ik_\parallel x}
  \left[
    \varphi_+(y)e^{ik_\perp y} + \varphi_-(y)e^{-ik_\perp y}
  \right].
\end{equation}
Within quasiclassical approximation, which is legitimate for
weak-coupling superconductors with $\Delta$ sufficiently smaller than
the Fermi energy, $\varphi_\alpha(y)$ obeys the Andreev equation
\begin{equation}
  \begin{pmatrix}
    -\alpha i\hbar v_\perp \partial_y & \Delta_\alpha(y,\phi) \\
    \Delta_\alpha^*(y,\phi) & \alpha i\hbar v_\perp \partial_y
  \end{pmatrix}
  \varphi_\alpha(y) = E \varphi_\alpha(y),
  \label{eq:Andreev}
\end{equation}
where $v_\perp = v_F \sin\phi$ is the $y$-component of the Fermi
velocity.  Equation \eqref{eq:Andreev} has a solution that decays
exponentially as $y$ increases, corresponding to the edge mode.  To
obtain the energy and wave function of the edge mode, we adopt a
uniform gap model, where $\Delta_\alpha(y,\phi)$ is replaced in all
space by the bulk gap function $\Delta e^{\alpha i m\phi}$.  Then,
the edge mode, which satisfies the specular boundary
condition $\psi(x,0) = 0$, has the energy
\cite{Huang_PRB2014,Tada_PRL2015,Wang_PRB2018,Sugiyama_JPSJ2020}
\begin{equation}
  E_{\rm edge} = -s_\phi \Delta \cos(m\phi),\
  s_\phi = {\rm sgn}[\sin(m\phi)].
\end{equation}
The normalized wave function is
\begin{equation}
  \psi_{\rm edge}(x,y)
  = e^{ik_\parallel x}
  \left(
    e^{ik_\perp y} - e^{-ik_\perp y}
  \right)\varphi(y),
\end{equation}
where
\begin{gather}
    \varphi(y) = \sqrt{\kappa_0}\,e^{-\kappa_0y}
  \left(
    \frac{1 + s_\phi}{2} \ket{1} + \frac{1 - s_\phi}{2} \ket{2}
  \right),
  \\
  \ket{1} = \frac{1}{\sqrt{2}}
  \begin{pmatrix} 1 \\ -1 \\ \end{pmatrix},\ 
  \ket{2} = \frac{1}{\sqrt{2}}
  \begin{pmatrix} 1 \\ 1 \\ \end{pmatrix},
\end{gather}
and $\kappa_0 = {\Delta |\sin(m\phi)|}/{\hbar v_\perp}$.

\begin{figure}
  \includegraphics[scale=1]{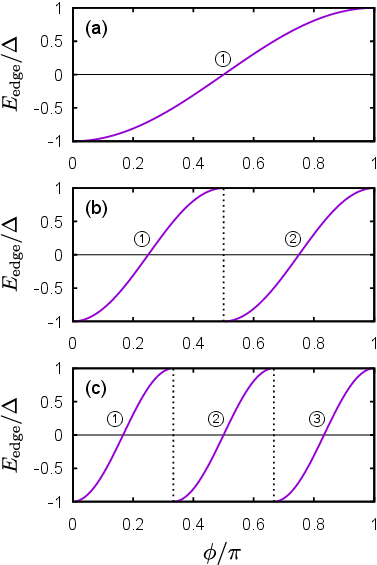}
  \caption{Energy of the edge mode as a function of $\phi/\pi$ in the
    specular limit for 2D chiral superconductors with (a) $p$-wave
    ($m=1$), (b) $d$-wave ($m=2$), and (c) $f$-wave ($m=3$) orbital
    symmetries.}
  \label{fig:Eedge-phi}
\end{figure}

The edge mode consists of $m$ branches, which are divided into two
groups: $s_\phi = +1$ and $s_\phi = -1$.
In Fig.\ \ref{fig:Eedge-phi}, we plot the $\phi$ dependence of
$E_{\rm edge}$ and number the branches \textcircled{\scriptsize 1},
\textcircled{\scriptsize 2}, \textcircled{\scriptsize 3}, $\cdots$ in
order of increasing $\phi$. The odd- and even-numbered branches are in
the $s_\phi = +1$ and $s_\phi = -1$ group, respectively. We
refer to the edge states in the $s_\phi = +1$ group as O-mode and those in
the $s_\phi = -1$ group as E-mode. As shown in Section\
\ref{sec:SDOS of the edge mode}, the O- and E-modes are coupled via
diffuse scattering when the surface is rough.

\section{Method}
\label{sec:Method}

We apply the quasiclassical theory of superconductivity to the above
model system.  The theory can be formulated in terms of the
quasiclassical Green's function $\hat{g}_\alpha(y)$, which is defined
as a $2 \times 2$ matrix in particle-hole space and obeys the
Eilenberger equation
\cite{Eilenberger_ZPhys1968}
\begin{gather}
  \alpha i \hbar v_\perp \partial_y \hat{g}_\alpha(y)
  = [\hat{g}_\alpha(y),\
  \hat{h}_\alpha(y,\phi,\varepsilon)],
  \label{eq:Eilen}
  \\
  \hat{h}_\alpha(y,\phi,\varepsilon) =
  \begin{pmatrix}
    \varepsilon & \Delta_\alpha(y,\phi) \\
    -\Delta_\alpha^*(y,\phi) & -\varepsilon \\
  \end{pmatrix},
  \label{eq:h-def}
\end{gather}
where $\varepsilon$ is a complex energy variable. Equation
\eqref{eq:Eilen} is supplemented by the normalization condition
$\hat{g}_\alpha^{\,2}(y) = -1$.  The normalized quasiclassical Green's
function in the bulk is given as
\begin{equation}
  \hat{g}_\alpha(\infty)
  = \frac{1}{\Omega}\hat{h}_\alpha(\infty,\phi,\varepsilon),
  \label{eq:g-bulk}
\end{equation}
where $\Omega = \sqrt{\Delta^2 - \varepsilon^2}\,$.

In this work, we are interested in the angle-resolved SDOS
\begin{equation}
  n(\phi,E) = {\rm Im}
  \left[
    \frac{1}{2}{\rm Tr}\sum_{\alpha=\pm}
    \frac{\hat{\rho}_3\hat{g}_\alpha^R(0)}{2}
  \right],
\end{equation}
where $\hat{\rho}_3$ is the third Pauli matrix in the particle-hole space
and $\hat{g}_\alpha^R(y)$ is the retarded quasiclassical Green's
function at $\varepsilon = E + i0^+$.  The gap function
$\Delta_\alpha(y,\phi)$ in Eq.\ \eqref{eq:h-def} is determined using the off-diagonal elements of the quasiclassical
Matsubara Green's function at $\varepsilon = i(2n+1)\pi k_B T$, where
$n = 0, \pm 1, \pm 2, \cdots$ (for details, see Ref.\
\onlinecite{Sugiyama_JPSJ2020}).

The quasiclassical Green's function is required to asymptotically
reach $\hat{g}_\alpha(\infty)$ as $y \to \infty$ and to satisfy a
surface boundary condition at $y = 0$.  When the surface is specular,
$\hat{g}_\alpha(0)$ satisfies
\begin{equation}
  \hat{g}_+(0) = \hat{g}_-(0) \equiv \hat{g}_S^{}.
  \label{eq:bcon-specular}
\end{equation}
This boundary condition implies that the quasiclassical propagator
on the specular reflection trajectory is continuous at the surface.
When the surface is rough, this property is lost due to
diffuse scattering. According to the random S-matrix theory, the
boundary condition is modified as
\cite{Nagato_JLTP1996,Nagato_JLTP1998,Nagai_JPSJ2008,Miyawaki_PRB2018}
\begin{equation}
  \hat{g}_+(0) =
  \frac{1+i\hat{\gamma}}{1-i\hat{\gamma}}\,
  \hat{g}_-(0)
  \frac{1-i\hat{\gamma}}{1+i\hat{\gamma}}.
  \label{eq:bcon-rough}
\end{equation}
The matrix $\hat{\gamma}$ carries information on diffuse scattering
effects and is self-consistently determined using
\begin{gather}
  \hat{\gamma} = 2W
  \left<
    \frac{1}{\displaystyle \hat{g}_S^{\,-1} - \hat{\gamma}}
  \right>_\phi,
  \label{eq:gamma-W}
  \\
  \left<\cdots\right>_\phi
  = \frac{\sum_{k_\parallel}(\cdots)}{\sum_{k_\parallel}1}
  = \frac{1}{2}\int_{0}^{\pi}d\phi\,\sin\phi\, (\cdots),
\end{gather}
where the parameter $W$ takes a value from zero to unity. The case
$W=0$ corresponds to the specular limit and $W = 1$ to the diffuse
limit at which an incident quasiparticle is scattered
isotropically. The rough surface effect
can be 
parameterized by
specularity $R$, defined as the specular reflection probability for an
electron at the Fermi level in the normal state.  The relationship between $W$ and $R$ is given by $W = (1-\sqrt{R})/(1+\sqrt{R})^2$
\cite{Miyawaki_PRB2018}.  In the
intermediate regime between $W = 0$ ($R = 1$) and $W = 1$ ($R = 0$),
specular reflection and diffuse scattering occur with probability $R$
and $1-R$, respectively.

The quasiclassical Green's function that satisfies the normalization
and boundary conditions can be expressed as (see
Appendix \ref{appen:A})
\begin{align}
  \hat{g}_+(y)
  &= \frac{2i\hat{\rho}_3}{1-D(y)F(y)}
  \begin{pmatrix} 1 \\ D(y) \\ \end{pmatrix}
  \begin{pmatrix} 1 & F(y) \\ \end{pmatrix} - i,
  \label{eq:g-p-DF}
  \\
  \hat{g}_-(y)
  &= \frac{2i\hat{\rho}_3}{1-D(y)F(y)}
  \begin{pmatrix} 1 \\ F(y) \\ \end{pmatrix}
  \begin{pmatrix} 1 & D(y) \\ \end{pmatrix} - i,
  \label{eq:g-m-DF}
\end{align}
where $D(y)$ and $F(y)$ obey the Riccati-type differential equations
\begin{align}
  i\hbar v_\perp \partial_y D
  &= 2\varepsilon D - \Delta_+^*(y,\phi) - \Delta_+(y,\phi)D^2,
    \label{eq:D-Riccati}
  \\
  i\hbar v_\perp \partial_y F
  &= -2\varepsilon F + \Delta_-^*(y,\phi) + \Delta_-(y,\phi)F^2,
    \label{eq:F-Riccati}
\end{align}
along with the boundary conditions
\begin{gather}
  D(\infty)
  = \frac{\Delta e^{-im\phi}}{\varepsilon + i\Omega},
  \label{eq:bcon-D}
  \\
  \begin{pmatrix} 1 \\ F(0) \\ \end{pmatrix}
  \propto
  \hat{\rho}_3 \frac{1-i\hat{\gamma}}{1+i\hat{\gamma}}\hat{\rho}_3
  \begin{pmatrix} 1 \\ D(0) \\ \end{pmatrix}.
  \label{eq:bcon-F}
\end{gather}

We note that $F(0) = D(0)$ holds in the specular limit and 
the following expression for $\hat{g}_S^{}$ is obtained.
\begin{align}
  \hat{g}_S^{}
  = \hat{\rho}_3
  \left(
  G_1 \ket{1}\!\bra{1} + G_2 \ket{2}\!\bra{2}
  \right),
  \label{eq:gS-G1-G2}
\end{align}
where
\begin{equation}
  G_1 = i\,\frac{1-D(0)}{1+D(0)},\
  G_2 = i\,\frac{1+D(0)}{1-D(0)}.
\end{equation}
In the next section, we show that $G_1$ and $G_2$ have the O-
and E-mode poles, respectively.  The $\hat{\gamma}$ matrix can be
parameterized in a form similar to Eq.\ \eqref{eq:gS-G1-G2}, i.e.,
\begin{equation}
  \hat{\gamma}
  = \hat{\rho}_3
  \left(
    S_1 \ket{1}\!\bra{1} + S_2 \ket{2}\!\bra{2}
  \right).
  \label{eq:gamma-S}
\end{equation}
Substituting Eq.\ \eqref{eq:gamma-S} into Eq.\ \eqref{eq:gamma-W}, we
determine $S_{1,2}$ as
\begin{align}
  S_1 &= 2W
        \left\langle \frac{1}{G_1^{-1} - S_2} \right\rangle_\phi,
        \label{eq:sc-S1}\\
  S_2 &= 2W
        \left\langle \frac{1}{G_2^{-1} - S_1} \right\rangle_\phi.
        \label{eq:sc-S2}
\end{align}

For an arbitrary $W$, $F(0)$ is given as
\begin{equation}
  F(0)
  = \frac{\mathbb{S}_1\mathbb{S}_2D(0) + (\mathbb{S}_1 - \mathbb{S}_2)/2}
  {1  - (\mathbb{S}_1 - \mathbb{S}_2)D(0)/2},
  \label{eq:F0-rough}
\end{equation}
where
\begin{equation}
  \mathbb{S}_j = \frac{1 + iS_j}{1 - iS_j} \quad (j = 1,2).
\end{equation}
Equations \eqref{eq:F0-rough}, \eqref{eq:g-p-DF}, and
\eqref{eq:g-m-DF} leads to the following expression for SDOS.
\begin{gather}
  n(\phi,E) = n_1(\phi,E) + n_2(\phi,E),
  \label{eq:SDOS-n}
  \\
  n_j(\phi,E)
  = \frac{1}{2}{\rm Im}(\mathcal{G}_j^R)\ \ (j=1,2),
\end{gather}
where
\begin{align}
  \mathcal{G}_1
  &= i\,\frac{1 - \mathbb{S}_2D(0)}{1 + \mathbb{S}_2D(0)}
    = \frac{G_1 + S_2}{1 - S_2G_1},
    \label{eq:cG1-def}\\
  \mathcal{G}_2
  &= i\,\frac{1 + \mathbb{S}_1D(0)}{1 - \mathbb{S}_1D(0)}
    = \frac{G_2 + S_1}{1 - S_1G_2}.
    \label{eq:cG2-def}
\end{align}
The Green's functions $\mathcal{G}_1$ and $\mathcal{G}_2$ have the
O- and E-mode poles affected by diffuse scattering,
respectively.

\begin{figure*}
  \includegraphics[scale=0.9]{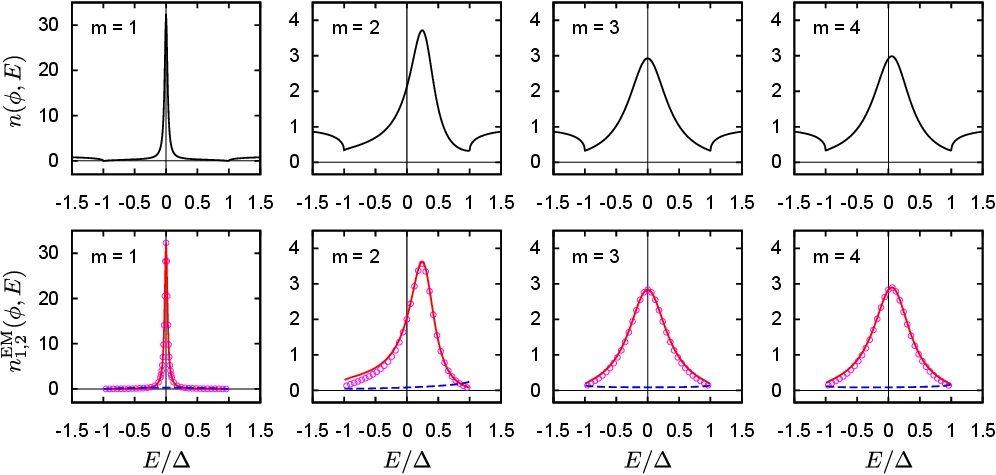}
  \caption{Angle-resolved SDOS at a rough surface with $W = 0.1$
    ($R = 0.5$) for $m = 1, 2, 3, 4$. The chosen angle is 
    $\phi = \pi/2m$, where a zero mode ($E_{\rm edge} = 0$) belonging
    to an O-mode branch appears in the specular limit.  The upper
    panels show the results of
    $n(\phi,E) = n_1(\phi,E) + n_2(\phi,E)$. The lower panels are
    plots from Eq.\ \eqref{eq:nj-def} for the subgap SDOS
    $n_1^{\rm EM}(\phi,E)$ (solid red lines) and
    $n_2^{\rm EM}(\phi,E)$ (dashed blue lines).  The open red circles
    correspond to the approximate formula \eqref{eq:nj-small-W} for
    the O-mode.}
  \label{fig:SDOS-full-and-small-W}
\end{figure*}

The energy dependence of angle-resolved SDOS for $m = 1,2,3,4$ and
$W = 0.1$ ($R = 0.5$) obtained from Eqs.\
\eqref{eq:SDOS-n}-\eqref{eq:cG2-def} within the uniform gap
approximation is shown in the upper panels of Fig.\ \ref{fig:SDOS-full-and-small-W}. The angle is chosen to be $\phi = \pi/2m$ as an
example.  The results illustrate the robustness of the edge mode to
diffuse scattering in the chiral $p$-wave superconductor ($m = 1$). For
the $p$-wave, SDOS of the edge mode is sharply peaked even
for the non-specular case ($W \neq 0$). In contrast, the broadening
of SDOS peak is significant for higher-order chiral states
($m \geq 2$). In the next section, we clarify the reason for the robustness of the edge
mode in the $p$-wave state.

\section{SDOS of the edge mode}
\label{sec:SDOS of the edge mode}

To elucidate the diffuse scattering effect on the edge mode, we
analyze SDOS under the uniform gap model, which allows analytical
calculations. The influence of the spatially varying self-consistent
gap function is discussed separately in Appendix \ref{appen:B}.

\subsection{Subgap SDOS}

Within the uniform gap model, $D(y)$ is a constant in space; thus,
$D(0) = D(\infty)$. Then, $G_{1,2}$ can be written as
\begin{equation}
  G_1 = -\frac{\Omega_1}{\varepsilon - E_1^0},\
  G_2 = -\frac{\Omega_2}{\varepsilon - E_2^0},
  \label{eq:G1-G2-ugap}
\end{equation}
where
\begin{align}
  \Omega_1 = \Omega + \Delta\sin(m\phi),\
  \Omega_2 = \Omega - \Delta\sin(m\phi),
\end{align}
and $E_1^0 = -\Delta \cos(m\phi)$ and $E_2^0 = \Delta \cos(m\phi)$ are
the energies of the O- and E-modes at the specular surface,
respectively.  From Eq.\ \eqref{eq:G1-G2-ugap}, we obtain the SDOS of
the edge modes at the specular surface ($W=0$) as
\begin{align}
  n^{\rm EM}_1(\phi,E)
  &= \frac{1+s_\phi}{2}\pi\Delta|\sin(m\phi)|\delta(E-E_1^0),
    \label{eq:n1-specular}\\
  n^{\rm EM}_2(\phi,E)
  &= \frac{1-s_\phi}{2}\pi\Delta|\sin(m\phi)|\delta(E-E_2^0).
    \label{eq:n2-specular}
\end{align}
Equations \eqref{eq:n1-specular} and \eqref{eq:n2-specular} correspond
to SDOS of the O- and E-modes, respectively.

To generalize these results to an arbitrary $W$, we
substitute Eq.\ \eqref{eq:G1-G2-ugap} into Eqs.\ \eqref{eq:cG1-def}
and \eqref{eq:cG2-def}. Then, we obtain
\begin{align}
  \mathcal{G}_1
  &= -\frac{\Omega_1 - (\varepsilon - E_1^0)S_2}
    {\varepsilon - E_1^0 + \Omega_1S_2},
    \label{eq:cG1-ugap}\\
  \mathcal{G}_2
  &= -\frac{\Omega_2 - (\varepsilon - E_2^0)S_1}
    {\varepsilon - E_2^0 + \Omega_2S_1}.
    \label{eq:cG2-ugap}
\end{align}
Taking the imaginary part of $\mathcal{G}_{1,2}^R$ for $|E| < \Delta$,
we get
\begin{equation}
  n_{j=1,2}^{\rm EM}(\phi,E)
  = -\frac{\Sigma_j''}{2\Omega_j}
  \frac{\Omega_j^2 + (E-E_j^0)^2}
  {(E-E_j^0-\Sigma_j')^2 + \Sigma_j''{}^2}.
  \label{eq:nj-def}
\end{equation}
Here, we introduce a surface self-energy defined as
\begin{equation}
  \Sigma_{1,2} = -\Omega_{1,2}S_{2,1}
\end{equation}
and denote the real and imaginary parts of the retarded self-energy as
$\Sigma_{1,2}'$ and $\Sigma_{1,2}''$, respectively. Note that
$\Sigma_{1,2}$ is proportional to $S_{2,1}$ (not $S_{1,2}$). This
implies that diffuse scattering between O- and E-modes is critical in the rough surface effect on SDOS (see the next
subsection).

\subsection{Lifetime of the edge mode}

When the surface is specular, the edge modes with energies $E_{1,2}^0$
have an infinite lifetime.
Hence, their SDOS are expressed as
delta-functions. In contrast, at rough surfaces, SDOS is broadened due to
diffuse scattering. In this subsection, we discuss SDOS in the case
$W \ll 1$, where the broadening is not large enough to obscure the well-defined edge modes.

Let the energies of the O- and E-modes at $W \neq 0$ be $E_1$ and
$E_2$, respectively. These energies can be defined as a solution of
\begin{equation}
  E - E_j^0 - \Sigma_j' = 0\ \ (j = 1,2).
\end{equation}
We expand the left-hand side around $E=E_j$ and introduce a
renormalization factor
\begin{equation}
  Z_j = \frac{1}{1-(\partial\Sigma_j'/\partial E)_{E=E_j}}.
\end{equation}
Then, we obtain
\begin{equation}
  E - E_j^0 - \Sigma_j' = (E-E_j)/Z_j.
  \label{eq:E-Z-rel}
\end{equation}
Substituting Eq.\ \eqref{eq:E-Z-rel} into Eq.\ \eqref{eq:nj-def}
yields
\begin{equation}
  n_j^{\rm EM}(\phi,E)
  = -\frac{Z_j^2\Sigma_j''}{2\Omega_j}
  \frac{\Omega_j^2 + (E-E_j^0)^2}
  {(E-E_j)^2 + (Z_j\Sigma_j'')^2}.
\end{equation}
In the numerator, $(E-E_j^0)^2 \sim (E_j-E_j^0)^2$ is
$O(\Sigma_j'{}^2)$ and can be neglected when $W \ll 1$. Thus, we arrive
at the following SDOS formula.
\begin{equation}
  n_j^{\rm EM}(\phi,E)
  = \frac{Z_j\Omega_j\Gamma_j/4}{(E-E_j)^2 + (\Gamma_j/2)^2},
  \label{eq:nj-small-W}
\end{equation}
where
\begin{equation}
  \Gamma_j = {\hbar}/{\tau_j} = -2Z_j\Sigma_j''
\end{equation}
is the inverse lifetime of the O-mode ($j=1$) and E-mode ($j=2$).

The
SDOS for the edge modes as a function of $E/\Delta$ is shown in the lower panels of Fig.\ \ref{fig:SDOS-full-and-small-W}. The solid red
and dashed blue lines, obtained using Eq.\
\eqref{eq:nj-def}, represent $n_1^{\rm EM}(\phi,E)$ and
$n_2^{\rm EM}(\phi,E)$, respectively. Since $\phi = \pi/2m$ is in the O-mode branch and
$W = 0.1$ is sufficiently small, the subgap SDOS is dominated by
$n_1^{\rm EM}(\phi,E)$. The open red circles are plots of the
approximate formula \eqref{eq:nj-small-W} for the O-mode,
demonstrating that Eq.\ \eqref{eq:nj-small-W} successfully reproduces SDOS of
the edge mode at small $W$.

\begin{figure}
  \includegraphics[scale=0.8]{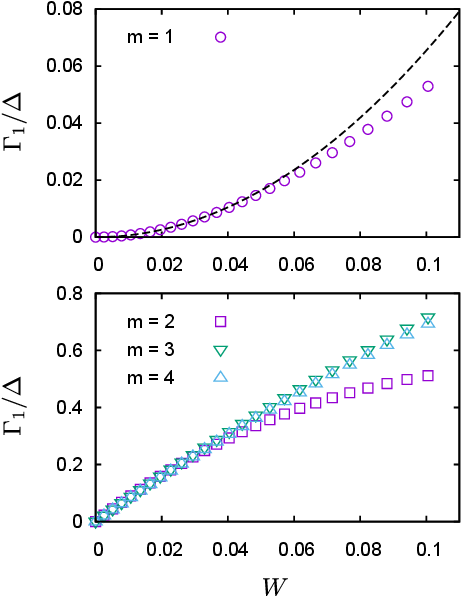}
  \caption{Inverse lifetime $\Gamma_1$ at $\phi = \pi/2m$ and
    $E = E_1$ as a function of $W$. The dashed line in the upper panel
    shows $W^2$ dependence of the low $W$
    data.}
  \label{fig:Gamma-W}
\end{figure}

In Fig.\ \ref{fig:Gamma-W}, the inverse lifetime $\Gamma_1$ at
$\phi = \pi/2m$ and $E = E_1$ is plotted as a function of $W <
0.1$. The $W$ dependence of $\Gamma_1$ is qualitatively different
between the chiral $p$-wave and other chiral states. As $W$ increases from zero,
$\Gamma_1$ in the chiral $p$-wave state increases proportionally to $W^2$,
while those in
the other chiral states increase linearly with $W$. Therefore, the inverse
lifetime in the chiral $p$-wave state is kept relatively small. Similar $W$ dependence of the inverse lifetime is observed for SDOS
calculated with the self-consistent gap function (see Appendix
\ref{appen:B}).

The $W$ difference of the inverse lifetime can be understood as
follows. The surface self-energy $\Sigma_1$ is proportional to
$S_2$. The imaginary part of $S_2$ of $O(W)$ is given by
\begin{equation}
  S_2''
  = 2W \left\langle {\rm Im}[G_2^R] \right\rangle_{\phi}
  = 4W \left\langle n_2^{\rm EM}(\phi,E)  \right\rangle_{\phi}.
\end{equation}
It follows that diffuse scattering between O- and E-modes is
responsible for the linear increase of the inverse lifetime with $W$.
The $W^2$ dependence of $\Gamma_1$ in the chiral $p$-wave state is because the edge mode has only a single branch. In this case,
the finite inverse lifetime is caused by diffuse scattering via
continuum states.

\subsection{SDOS in the diffuse limit}

\begin{figure}
  \includegraphics[scale=0.8]{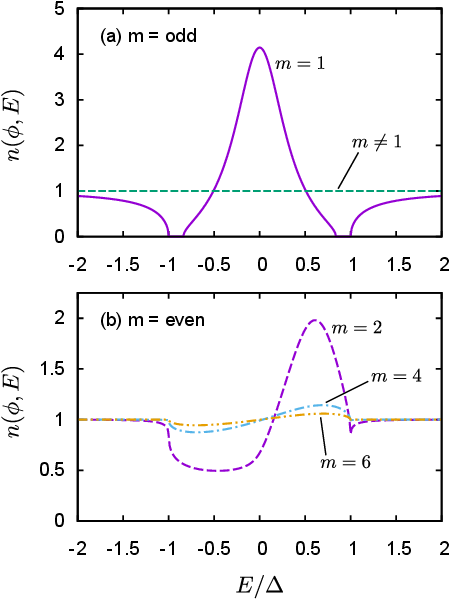}
  \caption{Angle-resolved SDOS $n(\phi, E)$ at $\phi = \pi/2m$ in the
    diffuse limit for (a) $m =$ odd and (b) $m =$ even. In the case of
    $m =$ odd, $n(\phi, E) = 1$ except when $m = 1$.}
  \label{fig:sdos-R0}
\end{figure}

Diffuse scattering causes the mixing of O- and E-modes. In the
case of $W \ll 1$, the mode
mixing is weak that subgap SDOS of the edge mode is described
by $n_1(\phi, E)$ or $n_2(\phi, E)$.  However, when $W \simeq 1$, the
classification of the edge mode into O- and E-modes is not useful
because of significant mode mixing. In particular, when $W = 1$, we
have $n_1(\phi, E) = n_2(\phi, E)$, as can be shown from the property
$\hat{\gamma}^2 = -1$ ($S_1S_2=-1$) of $\hat{\gamma}$ at $W = 1$
\cite{Nagai_JPSJ2008,Miyawaki_PRB2018}.

Figure \ref{fig:sdos-R0}(a) shows
$n(\phi,E) = n_1(\phi,E) + n_2(\phi,E)$ for $m =$ odd (spin triplet)
in the diffuse limit. The SDOS in the $p$-wave state has a peak
structure within the bulk gap, while those in non-$p$-wave triplet
states coincide with the normal-state value, $n(\phi,E) = 1$.  To
understand the origin of the difference between the
$p$- and non-$p$-wave triplet states, we assume that
$S_1 = S_2 = i$. This assumption leads to
$\mathcal{G}_1 = \mathcal{G}_2 = i$ [see Eqs.\ \eqref{eq:cG1-def} and
\eqref{eq:cG2-def}]; hence, $n(\phi,E) = 1$.  Substituting
$S_1=S_2=i$ into the right-hand side of Eqs.\ \eqref{eq:sc-S1} and
\eqref{eq:sc-S2}, we obtain
\begin{align}
  S_1 &= i(1 - \langle D(0) \rangle_\phi),\\
  S_2 &= i(1 + \langle D(0) \rangle_\phi).
\end{align}
It follows that the above assumption for $S_{1,2}$ is justified if
$\langle D(0) \rangle_\phi = 0$. Within the uniform gap model, we have
\begin{align}
  &(m = 1)\quad
    \langle D(0) \rangle_\phi
    = \frac{\pi}{4i}\frac{\Delta}{\varepsilon + i\Omega},
    \label{eq:ave-D-p-wave}\\
  &(m \neq 1)\quad
    \langle D(0) \rangle_\phi
    = \frac{1+(-1)^m}{2(1-m^2)}\frac{\Delta}{\varepsilon + i\Omega}.
    \label{eq:ave-D-non-p-wave}
\end{align}
Thus, $S_1=S_2=i$ is the self-consistent solution of $S_{1,2}$ for
non-$p$-wave triplet states in the diffuse limit. This observation
indicates that destructive interference due to diffuse scattering
and anisotropic gap of chiral states results in $n(\phi, E) = 1$.

In Fig.\ \ref{fig:sdos-R0}(b), we show SDOS for $m =$ even (spin
singlet) in the diffuse limit.  In the singlet states, SDOS does not
coincide with $n(\phi,E) = 1$ because of incomplete destructive
interference.  However, $n(\phi,E)$
approaches $n(\phi,E) = 1$ as $m$ increases, as
implied by Eq.\ \eqref{eq:ave-D-non-p-wave}, which shows that the
destructive interference becomes increasingly complete as $m$
increases.

The results confirm that SDOS in the diffuse limit is almost
structureless except for the chiral $p$- and $d$-wave
superconductors. This conclusion is not altered even when considering the
spatial variation of the self-consistent gap function (see Appendix
\ref{appen:B}).

\section{Conclusion}
\label{sec:Conclusion}

We studied the rough surface effect on the edge mode in 2D chiral
superconductors to elucidate how robust the edge mode is to diffuse
surface scattering. Specifically, we analyze the lifetime of the edge
mode extracted from the angle-resolved SDOS formulated using the
random S-matrix theory. We showed that when there are multiple
branches of the edge mode, inter-branch scattering occurs due to
diffuse scattering, considerably shortening the lifetime of the
edge mode. Multiple branches arise in the higher order
(non-$p$-wave) chiral superconductors. In contrast, in the
chiral $p$-wave state, only a single branch exists, resulting in a robust edge
mode.

In addition, we investigated the rough surface effect in the diffuse
limit, where lifetime analysis is less meaningful due to strong branch
mixing. The diffuse limit behavior of SDOS is understood qualitatively
in terms of destructive interference, such as in unconventional
superconductors with impurities.  The destructive interference renders
SDOS almost structureless in the chiral states other than the chiral
$p$- and $d$-wave states.  In particular, SDOS in the spin-triplet
non-$p$-wave chiral state coincides with that in the normal state. The
results confirm the findings of Suzuki and Asano
\cite{Suzuki_PRB2016}
for the edge current density.

Our theory is sufficiently general to cover edge modes other than those in 
chiral superconductors, such as include flat-band edge modes, which 
have received considerable attention in the context of non-chiral
$d$-wave superconductivity in cuprates
\cite{Hu_PRL1994,Tanaka_PRL1995,Nagato_PRB1995,Higashitani_JPSJ1997,
  Hakansson_NatPhys2015,Holmvall_NatCommun2018,Matsumoto_JPSJ1995,Yamada_JPSJ1996}.
As in the chiral case, the wave function for the flat band can be classified into two groups,
and the inter-branch scattering broadens the flat band.
In fact,
the flat-band broadening is significant in $d$-wave superconductors 
\cite{Matsumoto_JPSJ1995,Yamada_JPSJ1996} but not in $p$-wave
superconductors \cite{Bakurskiy_PRB2014,Nagato_JLTP1998}.
It should be noted,
however, that unlike the chiral case the SDOS broadening for the flat
band is accompanied by a peculiar
V-shaped structure at $E = 0$ \cite{Matsumoto_JPSJ1995}.
The origin of this structure remains an open question for future investigation.

\begin{acknowledgments}
  This work was supported in part by the JSPS KAKENHI Grant No.\ 22K03530.
\end{acknowledgments}

\appendix

\section{Quasiclassical Green's function}
\label{appen:A}

The quasiclassical Green's function can be constructed in terms of
two-component amplitude obeying the Andreev equation
\eqref{eq:Andreev} with $E$ replaced by the complex energy
$\varepsilon$.
The Andreev equation has damping ($\varphi_\alpha^d$) and growing
($\varphi_\alpha^g$) solutions with asymptotic behaviors
\begin{align}
  \varphi_{+}^{d}(y \to \infty)
  &\propto \hat{\rho}_1 \varphi_{-}^{d}(y \to \infty)
    \propto
    \begin{pmatrix}
      1 \\ D(\infty)
    \end{pmatrix}e^{-\kappa y},
    \label{appen:phi-d-inf}
  \\
  \varphi_{+}^{g}(y \to \infty)
  &\propto \hat{\rho}_1 \varphi_{-}^{g}(y \to \infty)
    \propto
    \begin{pmatrix}
      F(\infty) \\ 1
    \end{pmatrix}e^{\kappa y},
    \label{appen:phi-g-inf}
\end{align}
where
\begin{equation}
  D(\infty) = \frac{\Delta e^{-im\phi}}{\varepsilon + i\Omega},\
  F(\infty) = \frac{\Delta e^{im\phi}}{\varepsilon + i\Omega},\
  \kappa = \frac{\Omega}{\hbar v_\perp},
  \label{appen:D-F-inf}
\end{equation}
and $\hat{\rho}_1$ is the first Pauli matrix in the particle-hole space.
We introduce new amplitudes
\begin{equation}
  |y, l{\alpha}\} = \hat{\rho}_3\varphi_{\alpha}^{l}(y),\ \
  \{y, l{\alpha}| = [\hat{\rho}_1\varphi_{\alpha}^l(y)]^T,
\end{equation}
where $l=d,g$ and the superscript $T$ denotes transpose. These
amplitudes obey
\begin{align}
  \alpha i\hbar v_\perp \partial_y |y,l{\alpha}\}
  &= -\hat{h}_\alpha(y,\phi,\varepsilon)|y,l{\alpha}\},
    \label{appen:ket-like}
  \\
  \alpha i\hbar v_\perp \partial_y \{y,l{\alpha}|
  &= \{y,l{\alpha}| \hat{h}_\alpha(y,\phi,\varepsilon).
    \label{appen:bra-like}
\end{align}
It follows readily from Eqs.\ \eqref{appen:ket-like} and
\eqref{appen:bra-like} that the Wronskian
\begin{equation*}
  [\varphi_\alpha^g(y)]^T(-i\hat{\rho}_2)\varphi_\alpha^d(y)
  = \{y,g{\alpha}|y,d{\alpha}\} = -\{y,d{\alpha}|y,g{\alpha}\}
\end{equation*}
is a non-zero constant, and $|y,l{\alpha}\}\{y,l'{\alpha}|$ satisfies
the Eilenberger equation (Eq.\ \eqref{eq:Eilen}).  From the observation, we
arrive at the following expressions for $\hat{g}_\pm(y)$ that
reproduce the bulk form given in Eq.\ \eqref{eq:g-bulk}.
\begin{align}
  \hat{g}_+(y)
  &= \frac{2i}{\{y,g{+}|y,d{+}\}}|y,d{+}\}\{y,g{+}| - i,
    \label{appen:g-p-braket}\\
  \hat{g}_-(y)
  &= \frac{2i}{\{y,d{-}|y,g{-}\}}|y,g{-}\}\{y,d{-}| - i.
    \label{appen:g-m-braket}
\end{align}

Using Eqs.\ \eqref{appen:g-p-braket} and \eqref{appen:g-m-braket}, we 
convert the surface boundary condition \eqref{eq:bcon-rough} to
that for the Andreev amplitudes. The result can be written in the form
\begin{align}
  |0,g{-}\} &\propto \frac{1-i\hat{\gamma}}{1+i\hat{\gamma}}\, |0,d{+}\},
              \label{appen:bcon-phi-1}\\
  \{0,d{-}| &\propto \{0,g{+}|\, \frac{1+i\hat{\gamma}}{1-i\hat{\gamma}}.
              \label{appen:bcon-phi-2}
\end{align}
In the specular limit, these are reduced to
\begin{align}
  \varphi_{-}^g(0) &\propto \varphi_{+}^d(0),
                     \label{appen:bcon-phi-1-specular}\\
  \varphi_{+}^g(0) &\propto \varphi_{-}^d(0).
                     \label{appen:bcon-phi-2-specular}
\end{align}
Equations \eqref{appen:bcon-phi-1-specular} and
\eqref{appen:bcon-phi-2-specular} have a simple physical meaning. For
$E > \Delta$ ($\varepsilon = E + i0^+$), the growing and damping
solutions at $y = \infty$ correspond to incoming and outgoing waves,
respectively. Thus, in Eq.\ \eqref{appen:bcon-phi-1-specular}, the
general growing solution $\varphi_{-}^g(y)$ (with a damping
component mixed) represents such a scattering process that an
incoming electron-like quasiparticle is Andreev reflected as a
hole-like one. The simultaneously occurring normal reflection is
described by $\varphi_{+}^d(y)$ that represents an outgoing
electron-like quasiparticle. The amplitudes of the normal and Andreev
reflections at a specular surface are determined from Eq.\
\eqref{appen:bcon-phi-1-specular}.  Equation
\eqref{appen:bcon-phi-2-specular} determines these reflection
amplitudes for a hole-like wave incoming toward the surface.
Equations \eqref{appen:bcon-phi-1} and \eqref{appen:bcon-phi-2} are a
generalization of Eqs.\ \eqref{appen:bcon-phi-1-specular} and
\eqref{appen:bcon-phi-2-specular} to the rough surface case, respectively.

It is useful to note that the Andreev amplitude has the particle-hole
symmetry
$\varphi_{+}^l(y) \propto \hat{\rho}_1\varphi_{-}^l(y)$.
Thus, we introduce the parameterization
\begin{align}
  \varphi_{+}^d(y) &= u_{+}^d(y) \begin{pmatrix} 1 \\ D(y) \\ \end{pmatrix}
                    \propto \hat{\rho}_1\varphi_{-}^d(y),
                    \label{appen:D(y)-def}\\
  \varphi_{-}^g(y) &= u_{-}^g(y) \begin{pmatrix} 1 \\ F(y) \\ \end{pmatrix}
                    \propto \hat{\rho}_1\varphi_{+}^g(y).
                    \label{appen:F(y)-def}
\end{align}
Substituting these into Eqs. \eqref{appen:g-p-braket} and
\eqref{appen:g-m-braket} yields
\begin{align}
  \hat{g}_+(y)
  &= \frac{2i\hat{\rho}_3}{1-D(y)F(y)}
  \begin{pmatrix} 1 \\ D(y) \\ \end{pmatrix}
  \begin{pmatrix} 1 & F(y) \\ \end{pmatrix} - i.
  \label{appen:g-p-DF}
  \\
  \hat{g}_-(y)
  &= \frac{2i\hat{\rho}_3}{1-D(y)F(y)}
  \begin{pmatrix} 1 \\ F(y) \\ \end{pmatrix}
  \begin{pmatrix} 1 & D(y) \\ \end{pmatrix} - i.
  \label{appen:g-m-DF}
\end{align}
From Eq. \eqref{appen:bcon-phi-1}, we obtain
\begin{equation}
  \begin{pmatrix} 1 \\ F(0) \\ \end{pmatrix}
  \propto
  \hat{\rho}_3 \frac{1-i\hat{\gamma}}{1+i\hat{\gamma}}\hat{\rho}_3
  \begin{pmatrix} 1 \\ D(0) \\ \end{pmatrix}.
  \label{appen:bcon-F}
\end{equation}
Equation \eqref{appen:bcon-phi-2} leads to the same result as Eq.\
\eqref{appen:bcon-F} because of particle-hole symmetry.

Functions $D(y)$ and $F(y)$ obey the Riccati-type differential
equations
\begin{align}
  i\hbar v_\perp \partial_y D
  &= 2\varepsilon D - \Delta_+^*(y,\phi) - \Delta_+(y,\phi)D^2,
    \label{appen:D-Riccati}
  \\
  i\hbar v_\perp \partial_y F
  &= -2\varepsilon F + \Delta_-^*(y,\phi) + \Delta_-(y,\phi)F^2.
    \label{appen:F-Riccati}
\end{align}
The functions $D(y)$ and $F(y)$ can be obtained as follows:
First, we integrate Eq.\ \eqref{appen:D-Riccati} from $y = \infty$ with
the initial value $D(\infty)$ given in Eq.\ \eqref{appen:D-F-inf}. Next,
we substitute the obtained $D(0)$ into Eq. \eqref{appen:bcon-F} to find
$F(0)$. Finally, Eq.\ \eqref{appen:F-Riccati} is integrated from $y=0$.

\section{SDOS under the self-consistent gap function: Comparison to
  the uniform gap model}
\label{appen:B}

The numerical results of SDOS presented in the main text are obtained
under the uniform gap model. Here, we discuss the influence of the
spatially varying self-consistent gap function on SDOS.

\begin{figure}
  \includegraphics[scale=1]{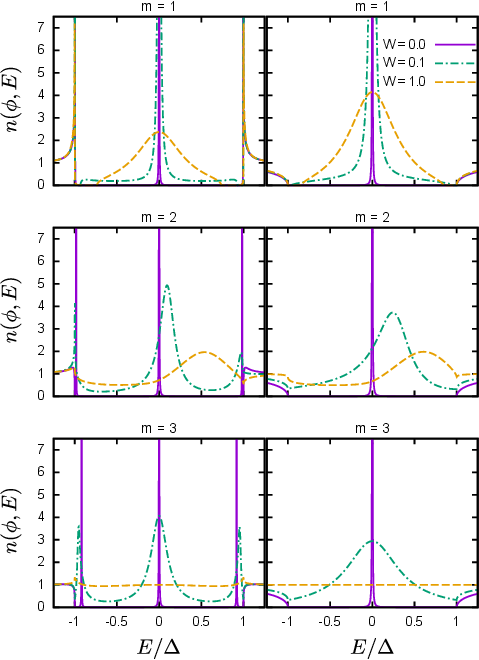}
  \caption{Angle-resolved SDOS at $\phi = \pi/2m$ for $m = 1,2,3$ and
    $W = 0.0, 0.1, 1.0$.  The left panels are the results computed
    with the self-consistent gap function.  The right panels are those
    under the uniform gap model.}
  \label{fig:sdos-comp}
\end{figure}

The left (right) panels of Fig.\ \ref{fig:sdos-comp} show the
angle-resolved SDOS at $\phi = \pi/2m$ computed with the
self-consistent (uniform) gap function. When the surface is specular
(solid violet lines), the self-consistent SDOS has sharp peaks
corresponding to edge modes of two: the gapless mode,
which is the focus of this paper, and
high-energy modes with $|E|$ close to the bulk gap edge
\cite{Sugiyama_JPSJ2020}. The peaks of
the high-energy modes are absent in SDOS under the uniform gap model.
This is because the high-energy modes appear due to the suppression of
$\Delta^a(y,\phi)$ at the surface (see Fig.\ \ref{fig:Delta-y}) and
are bound states trapped by the potential valley formed between the
surface and suppressed $\Delta^a(y,\phi)$
\cite{Sugiyama_JPSJ2020}.

Note that quasiparticles specularly scattered at the surface undergo a
difference in sign of $\Delta^a(y,\phi)$ between the incoming and
outgoing trajectories. Such scattering processes are known to 
suppress $\Delta^a(y,\phi)$ at the surface
\cite{Nagato_JLTP1998}.
In the chiral $p$-wave state, quasiparticles
can experience the sign change of $\Delta^a(y,\phi)$ in any scattering channel.
Thus, $\Delta^a(y,\phi)$ is strongly suppressed in the
specular and diffuse limits. In contrast, in
the higher-order chiral states, there are scattering channels in which
$\Delta^a(y,\phi)$ maintains its sign, weakening the suppression 
when the surface is rough. Therefore, the uniform gap model 
fairly agrees with the self-consistent SDOS for $W=0$ (dashed
yellow lines) in the higher-order chiral states.

\begin{figure}
  \includegraphics[scale=0.8]{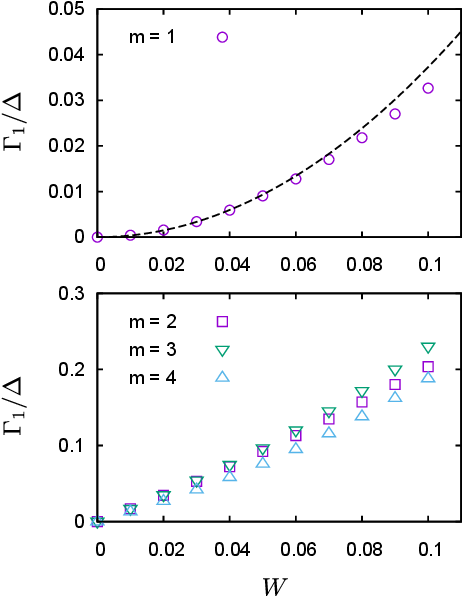}
  \caption{Inverse lifetime $\Gamma_1$ as a function of $W$ for the
    same parameters of $\phi$ and $E$ as those in Fig.\
    \ref{fig:Gamma-W}. The results are computed with the
    self-consistent gap function.}
  \label{fig:sc-Gamma-W}
\end{figure}

Except for the high-energy modes, the uniform gap model captures 
the qualitative features of the self-consistent SDOS. In particular,
the inverse lifetime extracted from the self-consistent SDOS by
Lorentzian fitting reproduces the $W$ dependence predicted by the
uniform gap model, i.e., a linear and quadratic increase from $W = 0$
in the chiral $p$-wave state and other chiral states,
respectively (Figs.\ \ref{fig:Gamma-W} and \ref{fig:sc-Gamma-W}).

%\bibliography{myrefs}

\end{document}